# On "Remarks on the spin-one Duffin-Kemmer-Petiau equation in the presence of nonminimal vector interactions in (3+1) dimensions"

H. Hassanabadi[1], Z. Molaee[2] and M. Ghominejad[2], S. Zarrinkamar[3]

[1] Physics Department, Shahrood University of Technology, P.O. Box 3619995161-316, Shahrood, Iran
[2] Physics Department, Semnan University, Semnan, Iran
[3] Department of Basic Sciences, Garmsar Branch, Islamic Azad University, Garmsar, Iran

In a very recent manuscript [1], Castro and Oliveira have commented on our recently published paper [2]. Their main criticism is that we have used an improper nonminimal interaction term. Regarding their work, we wish to mention two points.

1. We have started the paper based on the work of Kozak et al. [3] which has successfully discussed the deuteron-nucleus scattering.
2. The second point is that we have used in our calculations $\beta^0$ and not $\beta^v$ (which preserves the current conservation). Let us consider the following lines

The general form of the interaction is

$$U = S(r) + PS_v(r) + \beta^v V_v(r) + \beta^v PV_p(r), \qquad (1)$$

which, for an elastic scattering can be considered in the form

$$U = S(r) + PS_v(r) + \beta^0 V(r) + \beta^0 PV_P(r). \qquad (2)$$

The two Lorentz vectors may be written as $\beta^v$ and $P\beta^v$ by assuming rotational invariance and parity conservation. In our work, we have considered $\beta^0 PV_P(r)$ and therefore, the content of our paper is correct.

## References

[1] L. B. Castro and L. P. de Oliveira, "Remarks on the spin-one Duffin-Kemmer-Petiau equation in the presence of nonminimal vector interactions in (3+1) dimensions" arXiv:1403.6035v1 [hep-th].
[2] H. Hassanabadi, et al., "Spin-one DKP equation in the presence of coulomb and harmonic oscillator interactions in (1+3)-dimension," Advances in High Energy Physics, vol. 2012, Article ID 489641, 10 pages, 2012.
[3] R. E. Kozack, et al., "Spin-one Kemmer-Duffin-Petiau equation and intermediate-energy deuteron-nucleus scattering," Physical Review C, vol. 40, pp. 2181–2194, 1989.